\documentclass[a4paper]{jpconf}

\usepackage{amsmath,amsfonts,amssymb}
\usepackage{graphicx}

\begin{document}

\title{Cosmology with a Decaying Vacuum Energy Parametrization Derived from Quantum Mechanics\footnote{
Talk given at Seventh International Workshop DICE 2014: Spacetime -- Matter -- Quantum Mechanics \dots news on missing links, Castiglioncello (Tuscany, Italy), September 15--19, 2014 and submitted to the Proceedings of this conference.}}

\author{M~Szyd{\l}owski$^1$, A~Stachowski$^1$, K~Urbanowski$^2$}

\address{$^1$ Astronomical Observatory, Jagiellonian University,
Orla 171, 30-244 Krak{\'o}w, Poland}
\address{$^2$ Institute of Physics, University of Zielona G{\'o}ra}

\ead{marek.szydlowski@uj.edu.pl}

\begin{abstract}
Within the quantum mechanical treatment of the decay problem one finds that at late times $t$ the survival probability of an unstable state cannot have the form of an exponentially decreasing function of time $t$ but it has an inverse power-like form. This is a general property of unstable states following from basic principles of quantum theory. The consequence of this property is that in the case of false vacuum states the cosmological constant becomes dependent on time: $\Lambda - \Lambda_{\text{bare}}\equiv \Lambda(t) -\Lambda_{\text{bare}} \sim 1/t^{2}$. We construct the cosmological model with decaying vacuum energy density and matter for solving the cosmological constant problem and the coincidence problem. We show the equivalence of the proposed decaying false vacuum  cosmology with the $\Lambda(t)$ cosmologies (the $\Lambda(t)$CDM models). The cosmological implications of the model of decaying vacuum energy (dark energy) are discussed. We constrain the parameters of the model with decaying vacuum using astronomical data. For this aim we use the observation of distant supernovae of type Ia, measurements of $H(z)$, BAO, CMB and others. The model analyzed is in good agreement with observation data and explain a small value of the cosmological constant today.
\end{abstract}

\section{Introduction}
The nature and origin of an accelerating expansion of the Universe is the basic problem of modern cosmology. The most natural explanation of the acceleration of the Universe seems to be the cosmological constant parameter interpreted as vacuum energy. Although such an explanation is simple it brings a conundrum that the values of the cosmological constant required by quantum theory ($\sim 10^{71}$ GeV${}^{4}$) and obtained from type Ia supernovae observations ($\rho_{\Lambda} = \frac{\Lambda c^2}{8 \pi G} \sim 10^{-47}$ GeV${}^4$) differ about 100 orders of magnitude \cite{Weinberg:1988cp}. This problem is called the cosmological problem.

One approach toward the solving this problem is to consider the cosmological model with the time varying cosmological constant $\Lambda(t)$, where $t$ is the cosmological time. The parametrization for decaying vacuum is usually taken by hand to the cosmological model \cite{Alcaniz:2005dg,Wang:2004cp,Bessada:2013maa,Graef:2013iia}. Our idea is to derive the parametrization of decaying vacuum energy directly from the first principle, namely from the quantum mechanics, then to construct the cosmological model and to test it statistically using astronomical data. Therefore our approach to decaying vacuum cosmology is not purely phenomenological and is motivated by the fundamental theory of quantum mechanics.

Decaying false vacuum states from the point of view of the quantum theory of unstable states evolve in time. From basic principles of quantum theory it is known that the amplitude $A(t)$, and thus the decay law ${\cal P}_{M}(t) = |A(t)|^{2}$ of the unstable state $|M\rangle$, are completely determined by the density of the energy distribution function $\omega({ E})$ for the system in this state: $ A(t) = \int_{E_{min}}^{\infty} \omega({ E})\, \exp\, [-\frac{i}{\hbar}\,{ E}\,t]\,d{ E}$, where $\omega({E}) \geq 0$  and $\omega ({ E}) = 0$ for $E < E_{\text{min}}$. From this last condition and from the Paley--Wiener theorem it follows that there must be $|A(t)| \; \geq \; A_{1}\,\exp[- A_{2} \,t^{q}]$, for $|t| \rightarrow \infty$. Here $A_{1} > 0,\,A_{2}> 0$ and $ 0 < q < 1$. This means that the decay law ${\cal P}_{M}(t)$ of unstable states decaying in the vacuum can not be described by an exponential function of time $t$ if time $t$ is suitably long, $t \rightarrow \infty$, and that for these lengths of time ${\cal P}_{M}(t)$ tends to zero as $t \rightarrow \infty$  more slowly than any exponential function of $t$. It appears that these deviations from the exponential decay law at long times affect the energy of the unstable state and its decay rate at this time region and thus they affect the energy of the unstable false vacuum states at these times. It is shown in \cite{Urbanowski:2012pka,Urbanowski:2013tfa} that at transition times $t \sim T$, where $T$ denotes time when contributions of the exponential part of the survival probability and  of its  late time non--exponential part are equal, the instantaneous energy of the false vacuum states fluctuates and at late times, much latter then transition times, $t \gg T$,  it tends to the energy of the true vacuum state as $1/t^{2}$ for $t \to \infty$. The asymptotically late time behavior of the energy of the system in the false vacuum state is given by the following relation
\begin{equation}
{ E}^{\text{false}}_{0}(t) \simeq E^{\text{true}}_{0} \pm \frac{\alpha^2}{t^{2}} \cdots , \quad \text{for } t \gg T.
\label{E-false-infty-1}
\end{equation}
This means that in the case of such false vacuum states the cosmological constant becomes time depending. The standard relation is $\rho_{0}^{\text{true}} \equiv \rho_{\text{bare}} = \frac{\Lambda_{\text{bare}}}{8 \pi G}$. So the fluctuations of $\rho^{\text{false}}_{0}(t) =E^{\text{false}}_{0}(t)/{\bf V}$ at the transition times region and the asymptotically late behaviour of $\rho^{\text{false}}_{0}(t)$ at $t \gg T$ mean that identical behaviour of $\Lambda$ have to be observed at these times.

There are two possible scenarios. In the cosmological terminology if the universe is in a false vacuum state then the cosmological constant is time-dependent in the following way
\begin{equation}
\Lambda(t) = \Lambda_{\text{bare}} + \frac{3\beta}{t^{2}}, \quad \text{for } t \gg T.
\label{rho-false-infty-2a}
\end{equation}

In the model with decaying dark energy and matter, as in most cosmological models, we act on assumptions that the Universe is homogeneous and isotropic at the large scale (the cosmological principle), the structure and evolution of the Universe is governed by the Einstein theory of relativity, and the source of gravity is described by energy-momentum tensor for matter and decaying vacuum. It is also assumed for simplicity that the Universe is flat.

Then cosmological evolution is determined by the scale factor $a$ as a function of the universal cosmological time $t$, which satisfies the Friedmann equation
\begin{equation} \label{eq:1}
3 \frac{\dot{a}^{2}}{a^{2}} = \rho_{\text{m}} + \rho_{\text{vac}} = \rho_{\text{m}}+ \Lambda_{\text{bare}} + \frac{3\beta}{t^2}
\end{equation}
where matter density fulfils the conservation condition
\begin{equation}
\bar{T}^{\alpha \beta}_{\quad ;\beta} = T^{\alpha \beta}_{\text{m}} + \Lambda(t) g^{\alpha \beta}= 0.
\end{equation}
This condition for the flat homogeneous and isotropic Universe has the form
\begin{equation} \label{eq:5}
\dot{\rho}_{\text{m}} + 3 H \rho_{\text{m}} = - \dot{\Lambda}(t)
\end{equation}

In the case of the $\Lambda$CDM model ($\beta = 0$) it would be useful to rewrite (\ref{eq:1}) to the new form after substitution $a^3 = x^2$
\begin{equation} \label{eq:2}
\dot{x}^{2} = \frac{3}{4} \rho_{\text{m},0} + \frac{3}{4} \Lambda_{\text{bare}} x^2.
\end{equation}

In cosmology we use instead of density of fluids dimensionless parameters called density parameters
\[
\Omega_{i,0} = \frac{\rho_{i,0}}{3H_0^2}
\]
where $\rho_i$ is energy density of the fluid $i$, $H_0$ is the Hubble function $H = \dot{a}/a$ and the index $0$ denotes the present epoch.

Therefore eq.~(\ref{eq:2}) assumes the form
\begin{equation} \label{eq:7}
\left( \frac{dx}{d\tau} \right)^2 = \frac{9}{4} \Omega_{\text{m},0} + \frac{9}{4} \Omega_{\Lambda_{\text{bare}},0} x^2
\end{equation}
where $\Omega_{\Lambda_{\text{bare}}} = \frac{\Lambda}{3H_{0}^{2}}$, and $\tau \equiv |H_0 | t$.

Then we obtain the solution in the form
\begin{equation}
x(t) = \left( \frac{\Omega_{\text{m},0}}{\Omega_{\Lambda_{\text{bare}},0}} \right)^{1/2} \sinh \left( \frac{3}{2} \sqrt{\Omega_{\Lambda_{\text{bare}},0}}\ H_0 t \right).
\end{equation}
After dividing both sides of eq. (\ref{eq:1}) by $3H_{0}^{2}$, we obtain the relation
\begin{equation}
\left( \frac{H}{H_0} \right)^2 = \Omega_{\text{m},0} a^{-3} + \Omega_{\Lambda_{\text{bare}},0}.
\end{equation}
This relation can be rewritten in the terms of redshift $z$ as $1+z = a^{-1}$
\begin{equation}
\left( \frac{H}{H_0} \right)^2 = \Omega_{\text{m},0} (1+z)^{3} + \Omega_{\Lambda_{\text{bare}},0}.
\end{equation}
Today ($z=0$ and $H=H_0$) density parameters satisfy the constraint relation $\Omega_{\Lambda,0} = 1 - \Omega_{\text{m},0}$. This relation is fundamental for cosmography which analyzes the observational effects of photons propagation along the zero geodesics.

For the model with vacuum decaying
\begin{equation} \label{eq:9}
\left( \frac{H}{H_0} \right)^2 = \Omega_{\text{m}}(z) + \Omega_{\Lambda_{\text{bare}}} + \frac{\beta}{H_{0}^{2}} T(z)^{-2}
\end{equation}
where
\[
T(z) = - \int_{\infty}^{z} \frac{dz}{(1+z)H(z)}
\]
is the age of the Universe up to the redshift $z$.

For estimation of the model parameters we assume that
\[
H(z) = H_0 (1+z)^{\gamma}, \qquad \gamma = \text{const} > 0
\]
which gives
\[
T(z) = \frac{1}{\gamma H_0} (1+z)^{-\gamma} \propto H^{-1} = t_{H},
\]
where $t_{H}$ is a Hubble scale. Then
\begin{equation}
\left( \frac{H}{H_0} \right)^2 = \Omega_{\text{m}} + \Omega_{\Lambda_{\text{bare}}} + \frac{\beta \gamma^2}{H_{0}^{2}} H^{2}.
\end{equation}
The acceleration equation has the form
\begin{equation}
\dot{H} = \frac{\Lambda_{\text{bare}}}{2} - \delta H^2, \qquad \delta = \frac{3}{2} (1-\beta).
\end{equation}
After changing the variable $t \to a$ we obtain the acceleration equation in the form
\begin{equation}
\frac{dH}{da} \frac{da}{dt} = \frac{\Lambda_{\text{bare}}}{2} - \delta H^2
\end{equation}
or
\begin{equation} \label{eq:15}
\frac{dH}{da} = \frac{1}{Ha} \left( \frac{\Lambda_{\text{bare}}}{2} - \delta H^2 \right).
\end{equation}

The first integral of eq. (\ref{eq:15}) is
\begin{equation}
H^{2}(a) = \left( H_{0}^{2} - \frac{\Lambda_{\text{bare}}}{2\delta} \right) \left( \frac{a}{a_0} \right)^{-2\delta} + \frac{\Lambda_{\text{bare}}}{2\delta}.
\end{equation}
From this relation we obtain $H^2(z)$ formula which will be used in model parameter estimation
\begin{equation}
\left( \frac{H}{H_0} \right)^2 = \frac{3}{2\delta} \Omega_{\text{m}} (1+z)^{2\delta} + \frac{3}{2\delta} \Omega_{\Lambda_{\text{bare}}}
\end{equation}
where (further let write down $\Lambda_{\text{bare}} = \Lambda$)
\[
\Omega_{\text{m},0} +  \Omega_{\Lambda,0} = \frac{2}{3}\delta.
\]
When $\delta = 3/2$ the model reduces to the standard $\Lambda$CDM model.

Finally, we obtain the following model
\begin{equation}
\left( \frac{H}{H_0} \right)^2 = \frac{\Omega_{\text{m},0}}{\Omega_{\text{m},0}+\Omega_{\Lambda,0}} (1+z)^{3(\Omega_{\text{m},0}+\Omega_{\Lambda,0})} + \frac{\Omega_{\Lambda,0}}{\Omega_{\text{m},0}+\Omega_{\Lambda,0}}.
\label{eq:model}
\end{equation}

After the generalization of the corresponding substitution
$
a^{3(\Omega_{\text{m},0}+\Omega_{\Lambda,0})}=x^2,
$
we obtain the expression for the scale factor in following form
\[
a(t) = \left( \frac{\Omega_{\text{m},0}}{\Omega_{\Lambda,0}} \right)^{\frac{1}{3(\Omega_{\text{m},0}+\Omega_{\Lambda,0})}} \left[ \sinh\left(\frac{3}{2}\sqrt{\Omega_{\Lambda,0}}H_0 t \right) \right]^{\frac{2}{3(\Omega_{\text{m},0}+\Omega_{\Lambda,0})}}.
\]

\section{Data}

For the estimation of parameters of the model (\ref{eq:model}) we used the observations of SNIa data, BAO, CMB, measurements of $H(z)$, Alcock-Paczy{\'n}ski test.

First, we consider the SNIa data. The likelihood function is
\begin{equation}\label{like_sn}
\ln L_{\text{SNIa}} = -\frac{1}{2} \sum_{i=1}^{N}  \left (\frac{\mu_i^{\text{obs}}-\mu_i^{\text{th}}}{\sigma_i }\right)^2 ,
\end{equation}
where the summing is over the SNIa sample; the distance modulus $\mu^{\text{obs}}=m-M$ (where $m$ is the apparent magnitude and  $M$ is the absolute magnitude of SNIa stars) and $\mu^{\text{th}} = 5 \log_{10} D_L +25$ (where the luminosity distance is $D_L= c(1+z) \int_{0}^{z} \frac{d z'}{H(z')}$ and $\sigma$ is the uncertainties. The Union 2.1 sample of 580 supernovae was used as the data here \cite{Suzuki:2011hu}.

The BAO (baryon acoustic oscillation) data were taken from the Sloan Digital Sky Survey Release 7 (SDSS R7) dataset which consists of 893 319 galaxies \cite{Percival:2009xn}. The likelihood function is given by
\[
\ln L_{\text{BAO}} = \frac{\left( \frac{r_s(z_d)}{D_V(z)} - d(z) \right)^2}{\sigma^2}
\]
where $r_s(z_d)$ is the sound horizon at the drag epoch and $z=0.275$, $d(z)=0.1390$, $\sigma = 0.0037$ \cite{Eisenstein:1997ik}.

Planck observations of cosmic microwave background (CMB) radiation were also used \cite{Ade:2013zuv}. With addition information on lensing from the Planck and low-$\ell$ polarization from the WMAP (WP), we obtain the combined likelihood function in the form
\begin{equation}
\ln L_{\text{CMB}+\text{lensing}+\text{WP}} = - \frac{1}{2} \sum_{ij} (x_i^{\text{th}}-x_i^{\text{obs}}) \mathbb{C}^{-1} (x^{\text{th}}-x^{\text{obs}}),
\end{equation}
where $\mathbb{C}$ is the covariance matrix with the errors, $x$ is a vector of the acoustic scale $l_{A}$, the shift parameter $R$ and $\Omega_{b}h^2$ where
\begin{align}
l_A &= \frac{\pi}{r_s(z^{*})} c \int_{0}^{z^{*}} \frac{dz'}{H(z')} \\
R &= \sqrt{\Omega_{\text{m}} H_0^2} \int_{0}^{z^{*}} \frac{dz'}{H(z')}
\end{align}
where $z^{*}$ is the recombination redshift.

The Alcock-Paczynski test is the comparison of the radial and tangential size of an object, which is isotropic in the correct choice of model \cite{Alcock:1979mp,Lopez-Corredoira:2013lca}. The likelihood function is independent on the $H_0$ parameter and has the following form
\begin{equation}
\ln L_{AP} =  - \frac{1}{2} \sum_i \frac{\left( AP^{\text{th}}(z_i)-AP^{\text{obs}}(z_i) \right)^2}{\sigma^2}.
\end{equation}
where $AP(z)^{\text{th}} \equiv \frac{H(z)}{z} \int_{0}^{z} \frac{dz'}{H(z')}$ and $AP(z_i)^{\text{obs}}$ are observational data \cite{Sutter:2012tf,Blake:2011ep,Ross:2006me,Marinoni:2010yoa,daAngela:2005gk,Outram:2003ew,Anderson:2012sa,Paris:2012iw,Schneider:2010hm}.

At the end it is also valuable to add the constraints on the Hubble parameter, i.e. $H(z=0)\equiv H_0$. Data of $H(z)$ for samples of different galaxies were also used \cite{Simon:2004tf,Stern:2009ep,Moresco:2012jh}
\begin{equation}\label{hz}
  \ln L_{H(z)} = -\frac{1}{2} \sum_{i=1}^{N}  \left (\frac{H(z_i)^{\text{obs}}-H(z_i)^{\text{th}}}{\sigma_i }\right)^2.
\end{equation}

The final likelihood function for the observational Hubble function is
\begin{equation}
 L_{\text{tot}} = L_{\text{SNIa}} L_{\text{BAO}} L_{\text{CMB}+\text{lensing}+\text{WP}} L_{\text{AP}} L_{H(z)}.
\end{equation}

\section{Estimation of the model}

To estimate the model parameters we use our own code CosmoDarkBox implementing the Metropolis-Hastings algorithm \cite{Metropolis:1953es,Hastings:1970mc} and using the Pade approximants \cite{Aviles:2014rma} for the calculation of the likelihood function \cite{Hu:1995en,Eisenstein:1997ik}.

We use observation data of 580 supernovae of type Ia, selected subsets of the data points of Hubble function, the measurements of BAO from SDSS+2dSGRS. We also use data for the application of the Alcock-Paczynski test -- 18 observational points. At last, we estimated model parameters with CMB data from Planck, low-$\ell$ polarization from WMAP and lensing from Planck.

\begin{figure}
\begin{center}
\includegraphics[scale=0.25]{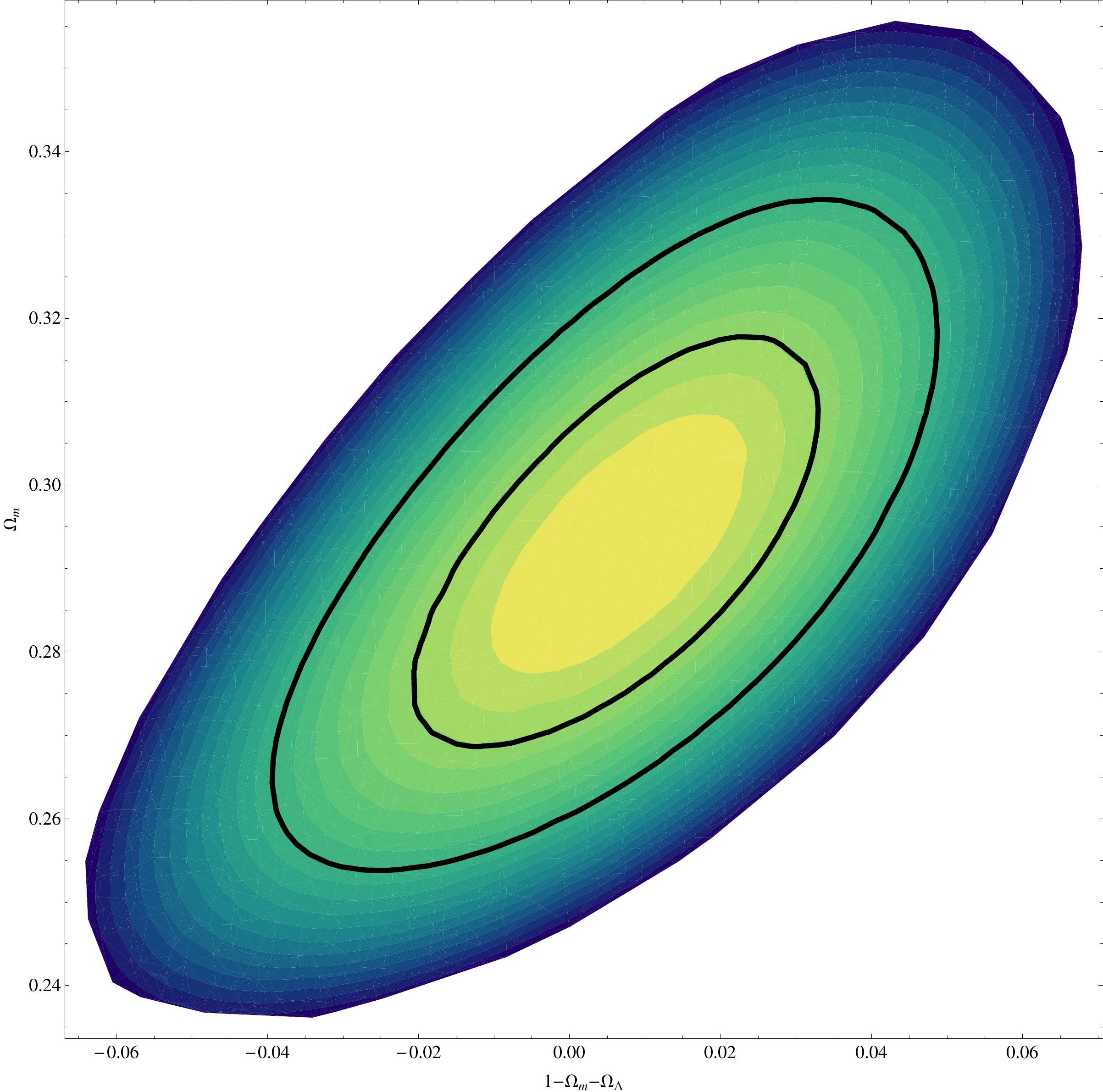}
\caption{\label{fig:1} The likelihood function of two model parameters ($1-\Omega_{\text{m}}-\Omega_{\Lambda}, \Omega_{\text{m}}$) with the marked $68\%$ and $95\%$ confidence levels. The value of Hubble constant is estimated from the data as best fit value $H_0=68.22$ km/(s Mpc) and then the figure is obtained for this value.}
\end{center}
\end{figure}

\begin{figure}
\begin{center}
\includegraphics[scale=0.25]{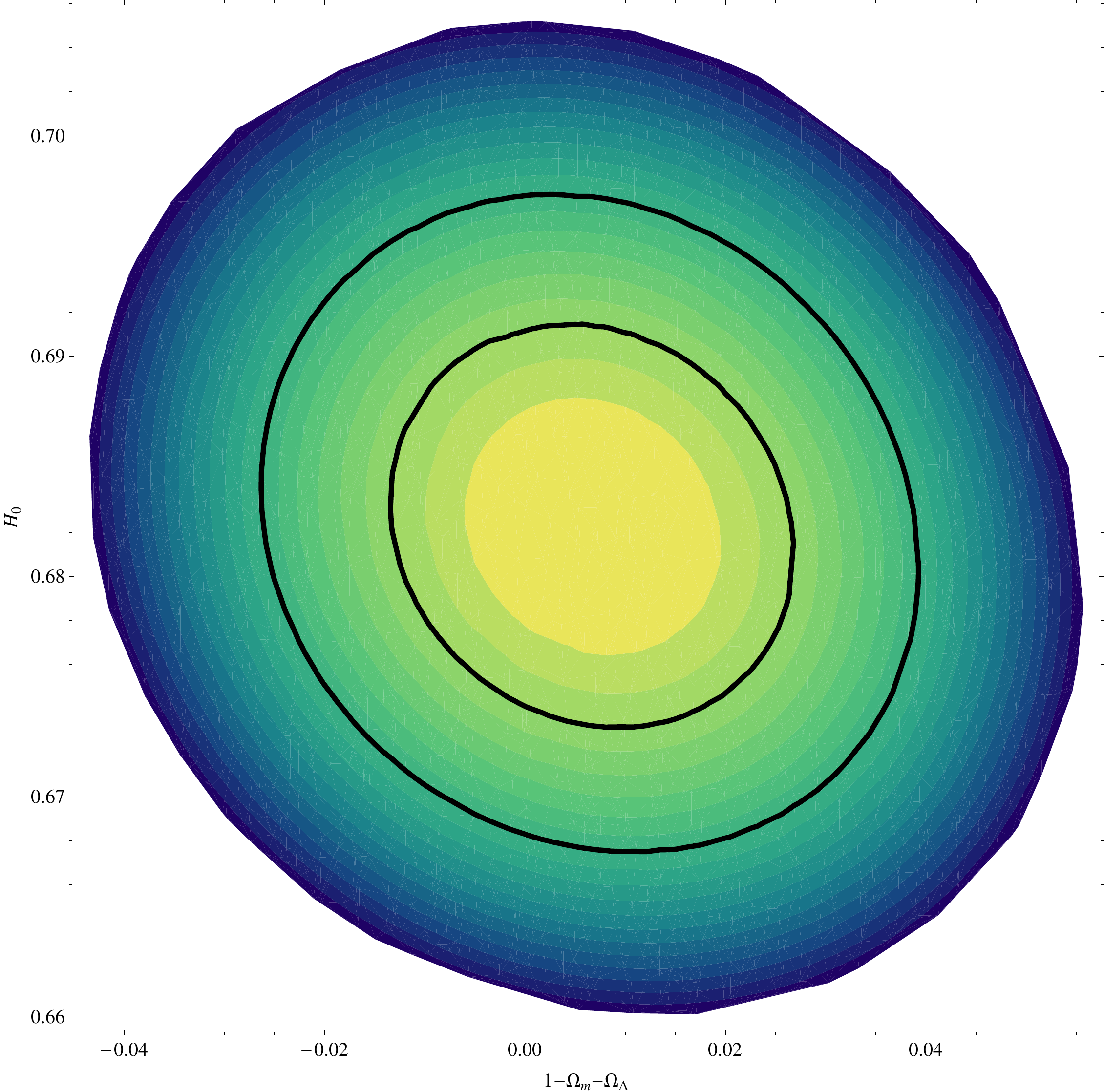}
\caption{\label{fig:2} The likelihood function of two model parameters ($1-\Omega_{\text{m}}-\Omega_{\Lambda}, H_0$) with the marked $68\%$ and $95\%$  confidence levels. The value of $\Omega_{\text{m}}$ is estimated from the data as best fit value 0.2926 and then figure is obtained for this value. Note that the effect of decaying vacuum appears as the value of the parameter $1-\Omega_{\text{m}}-\Omega_{\Lambda}$ is different from zero.}
\end{center}
\end{figure}

\begin{figure}
\begin{center}
\includegraphics[scale=0.4]{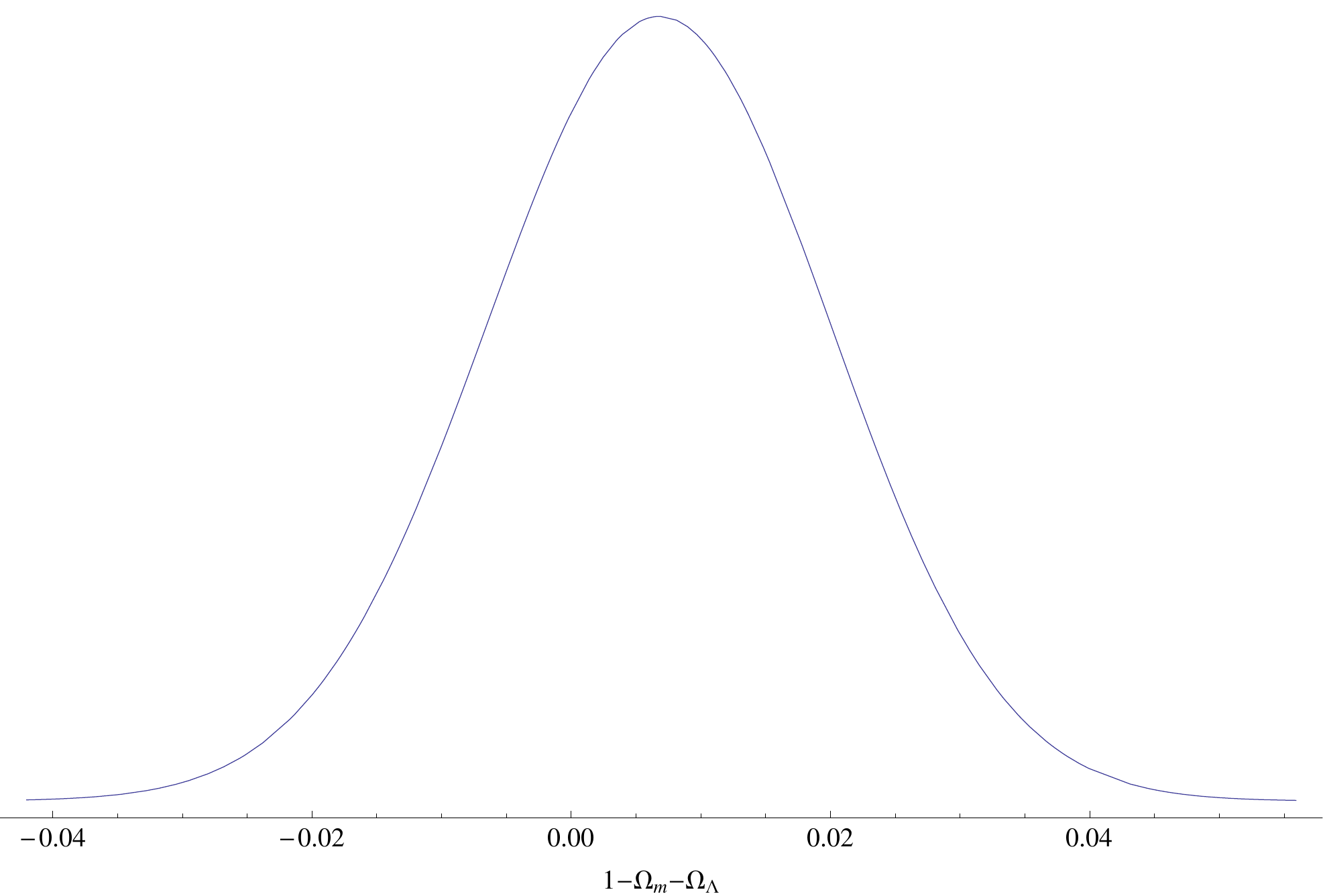}
\caption{\label{fig:3} Diagram of PDF for parameter $1-\Omega_{\text{m}}-\Omega_{\Lambda}$ obtained as an intersection of the likelihood function. Two planes of intersection likelihood function are $H_0=68.22$ km/(s Mpc) and $\Omega_{\text{m}}=0.2926$. The planes of intersection are constructed from the best fitting value of the model parameters. The maximum of PDF is reached for $1-\Omega_{\text{m}}-\Omega_{\Lambda}=0.0068$.}
\end{center}
\end{figure}

\begin{figure}
\begin{center}
\includegraphics[scale=0.4]{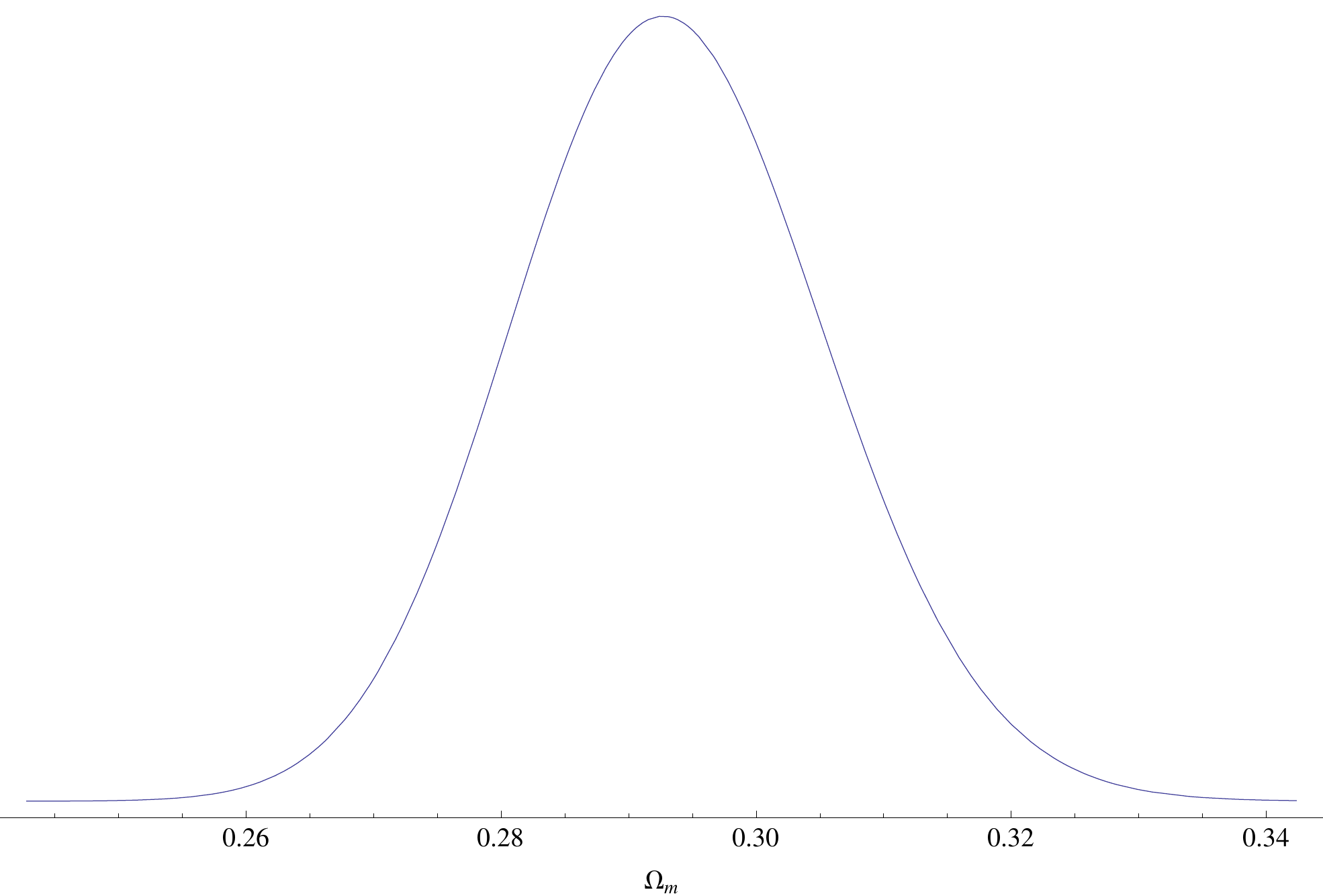}
\caption{\label{fig:4} Diagram of PDF for parameter $\Omega_{\text{m}}$ obtained as an intersection of the likelihood function. Two planes of intersection likelihood function are $H_0=68.22$ km/(s Mpc) and $1-\Omega_{\text{m}}-\Omega_{\Lambda}=0.0068$. The planes of intersection are constructed from the best fitting value of the model parameters. The maximum of PDF is reached for $\Omega_{\text{m}}=0.2926$.}
\end{center}
\end{figure}

The results of statistical analysis are represented in Figures~\ref{fig:1}--\ref{fig:4}. The values of estimated parameter are completed in Table 1. In Figures \ref{fig:1} and \ref{fig:2} it is shown the likelihood function with $68\%$ and $95\%$ confidence levels projection on the ($1-\Omega_{\text{m}}-\Omega_{\Lambda}, \Omega_{\text{m}}$) plane and the ($1-\Omega_{\text{m}}-\Omega_{\Lambda}, H_0$) plane, respectively. In figures \ref{fig:3} and \ref{fig:4} the intersections with respect to fixed $H_0$ and $\Omega_{\Lambda}$ for $\Omega_{m}$; and fixed $H_0$ and $\Omega_{m}$ for $\Omega_{\Lambda}$ have been presented, respectively.

We obtain, that the effect of decaying vacuum is rather small and errors one order higher. Therefore the effect cannot be detected from available observations data. Second, the errors for parameter $1-\Omega_{m}-\Omega_{\Lambda}$ are the same order like $\Omega_{m}, \Omega_{\Lambda}$ and we are looking for small parameter close to zero.

\begin{table}
\caption{\label{tab:1} The values of estimated parameters -- best fitted values of parameters with errors. We consider two cases for errors estimation, first $H_0$ is assumed as 68.22 km/(s Mpc), then $\Omega_{m}$ is assumed as the best fit value.}
\begin{center}
\begin{tabular}{lllllll}
\br
& parameter & best fit & $68\%$ CL & $95\%$ CL  \\
\mr
Case 1 -- assumed $H_0=68.22$ & $1-\Omega_{\text{m}}-\Omega_{\Lambda}$ & 0.0068 & ${}^{+0.0261}_{-0.0276}$ & ${}^{+0.0419}_{-0.0461}$\\
& $\Omega_{\text{m}}$ & 0.2926 & ${}^{+0.0251}_{-0.0239}$ & ${}^{+0.0416}_{-0.0388}$ \\
\mr
Case 2 -- assumed $\Omega_{\text{m}}=0.2926$ & $1-\Omega_{\text{m}}-\Omega_{\Lambda}$ & 0.0068 & ${}^{+0.0200}_{-0.0203}$ & ${}^{+0.0324}_{-0.0331}$\\
& $H_0$ & 68.22 & ${}^{+0.92}_{-0.91}$ & ${}^{+1.50}_{-1.45}$\\
\br
\end{tabular}
\end{center}
\end{table}

\section{Conclusion}

\begin{enumerate}
\item The cosmological model with decaying vacuum explains while the value of cosmological constant is so small --- possible solution of the cosmological constant problem.
\item The model with decaying vacuum (cosmological constant) was tested using astronomical data (SNIa, BAO, CMB, $H(z)$).
\item The evidence of decaying dynamical vacuum effect for the current Universe is equivalent to the sum of $\Omega_{m,0} + \Omega_{\Lambda,0} \ne 1$.
For the standard cosmological model this sum is equal one.
\item The value of sum of $\Omega_m$ and $\Omega_{\Lambda}$ is close to 1 (the obtained value is $0.993$) --- the effect of decaying vacuum is very weak.
\item The cosmological models with decaying vacuum be can treated as an extension of the standard cosmological model. This model is in a good agreement with astronomical data, and it offers the solution of the cosmological constant conundrum.
\end{enumerate}

\ack
The work was supported by the grant NCN DEC-2013/09/B/ST2/03455.

\section*{References}
\providecommand{\newblock}{}

\end{document}